\documentclass[%
 reprint,
nofootinbib,
 amsmath,amssymb,
 aps,
 prd,
]{revtex4-2}

\usepackage{graphicx}
\usepackage{dcolumn}
\usepackage{bm}
\usepackage{hyperref}
\usepackage{cleveref}
\usepackage[mathlines]{lineno}
\usepackage[dvipsnames]{xcolor}
\usepackage{tabularx}
\usepackage{booktabs}
\usepackage{multirow}

\usepackage{aasmacros}
\usepackage{units}
\usepackage{xspace}

\usepackage{subfigure}
\usepackage{acronym}

\newcommand{\nball}[1]{$#1$\nobreakdash\discretionary{-}{-}{-}ball }
\newcommand{\nsphere}[1]{$#1$\nobreakdash\discretionary{-}{-}{-}sphere }
\newcommand{\ndimensional}[1]{$#1$\nobreakdash\discretionary{-}{-}{-}dimensional}
\newcommand{\latent}{\mathcal{Z}}
\newcommand{\physical}{\mathcal{X}}
\newcommand{\diff}{\textrm{d}}
\renewcommand{\vec}[1]{\boldsymbol{#1}}

\newcommand{\nessai}{{\sc Nessai}\xspace}
\newcommand{\bilby}{{\sc Bilby}\xspace}
\newcommand{\bilbypipe}{{\sc bilby\_pipe}\xspace}
\newcommand{\lalinference}{{\sc LALInference}\xspace}
\newcommand{\dynesty}{{\sc dynesty}\xspace}
\newcommand{\cpnest}{{\sc cpnest}\xspace}
\newcommand{\nflows}{{\sc nflows}\xspace}
\newcommand{\pytorch}{{\sc PyTorch}\xspace}
\newcommand{\corner}{{\sc corner}\xspace}
\newcommand{\matplotlib}{{\sc matplotlib}\xspace}
\newcommand{\seaborn}{{\sc seaborn}\xspace}
\newcommand{\numpy}{{\sc NumPy}\xspace}
\newcommand{\scipy}{{\sc SciPy}\xspace}
\newcommand{\pandas}{{\sc pandas}\xspace}
\newcommand{\python}{{\sc Python}\xspace}

\newcommand{\imrphenomp}{{\sc IMRPhenomPv2}\xspace}

\newcolumntype{P}{>{\tt}c}

\newcommand{\figwidth}{8.6cm}
\newcommand{\onehalffigwidth}{12.9cm}
\newcommand{\doublefigwidth}{17.2cm}

\begin{document}

\newcommand{\evaluationsratio}{2.07\xspace}
\newcommand{\evaluationsratiomarg}{1.34\xspace}
\newcommand{\nessaievaluations}{$5.04 \times 10^{6}$\xspace}
\newcommand{\nessaievaluationsmarg}{$7.22 \times 10^{6}$\xspace}
\newcommand{\dynestyevaluations}{$10.44 \times 10^{6}$\xspace}
\newcommand{\dynestyevaluationsmarg}{$9.67 \times 10^{6}$\xspace}

\newcommand{\npopulation}{263\xspace}
\newcommand{\trainingits}{87\xspace}

\newcommand{\timeratio}{2.32\xspace}
\newcommand{\timeratiomarg}{1.40\xspace}
\newcommand{\nessairuntime}{\#\xspace}
\newcommand{\dynestyruntime}{\#\xspace}
\newcommand{\populationtime}{40\%\xspace}
\newcommand{\trainingtime}{8\%\xspace}
\newcommand{\populationtimemarg}{42\%\xspace}
\newcommand{\trainingtimemarg}{5\%\xspace}

\newcommand{\nessaipvalue}{0.3394\xspace}
\newcommand{\nessaipvaluemarg}{0.6818\xspace}

\newcommand{\logzerror}{0.092\xspace}
\newcommand{\logzsigma}{0.11\xspace}

\newcommand{\multipoptime}{36\%\xspace}
\newcommand{\multitraintime}{9\%\xspace}
\newcommand{\multievaltime}{54\%\xspace}
\newcommand{\multievaltimemin}{9\%\xspace}

\newcommand{\cornersnr}{15.54}

\preprint{APS/123-QED}

\title{Nested Sampling with Normalising Flows for Gravitational-Wave Inference}

\author{Michael J. Williams}
\author{John Veitch}
\author{Chris Messenger}%
\affiliation{%
SUPA, School of Physics and Astronomy \\
 University of Glasgow \\
 Glasgow G12 8QQ, United Kingdom
}%


\date{\today}

\begin{abstract}

We present a novel method for sampling iso-likelihood contours in nested sampling using a type of machine learning algorithm known as normalising flows and incorporate it into our sampler \nessai. \nessai is designed for problems where computing the likelihood is computationally expensive and therefore the cost of training a normalising flow is offset by the overall reduction in the number of likelihood evaluations. We validate our sampler on 128 simulated gravitational wave signals from compact binary coalescence and show that it produces unbiased estimates of the system parameters.
Subsequently, we compare our results to those obtained with \dynesty and find good agreement between the computed log-evidences whilst requiring \evaluationsratio times fewer likelihood evaluations. We also highlight how the likelihood evaluation can be parallelised in \nessai without any modifications to the algorithm. Finally, we outline diagnostics included in \nessai and how these can be used to tune the sampler's settings.

\end{abstract}

\maketitle

\acrodef{CBC}[CBC]{compact binary coalescence}
\acrodef{GW}[GW]{gravitational wave}
\acrodef{iid}[i.i.d.]{independently and identically distributed}
\acrodef{MCMC}[MCMC]{Markov Chain Monte Carlo}
\acrodef{SNR}[SNR]{signal-to-noise ratio}

\section{Introduction}

Gravitational-wave astronomy has contributed to our understanding of physics and astrophysics from the atomic scale up to the scale of the Universe \cite{GBM:2017lvd,Abbott:2018exr,Abbott:2019yzh}. This is set to continue as the detectors of the LIGO-Virgo-Kagra (LVK) Collaboration \cite{TheLIGOScientific:2014jea,TheVirgo:2014hva,Akutsu:2020his} continue to improve in sensitivity \cite{Aasi:2013wya} and new detectors come online \cite{Iyer:2011indigo} increasing the potential of detecting previously unseen types of sources.

Prior to the recent third observing run there were 11 confirmed detections of \acp{GW} from \ac{CBC} \cite{LIGOScientific:2018mvr}, including the first multi-messenger detection \cite{GW170817discovery}. The first half of the last year-long observing run has resulted in a further 39 candidate detections \cite{GWTC2:2020}. Each of the candidate events requires extensive analysis to determine whether they are astrophysical in origin and consequently allow us to understand the astrophysics that produce such phenomena. The nature of these analyses and ever increasing number of candidates necessitates more efficient analysis techniques.

Our understanding of the sources that produce the detected \acp{GW} hinges on the ability to infer the parameters that describe them. This inference is carried out in a Bayesian framework \cite{Aasi:2013jjl,TheLIGOScientific:2016wfe} centred around Bayes' theorem which allows for prior knowledge to be updated with observations to obtain posterior distributions that describe the probability of some parameters $\vec{\theta}$ given the observed data $d$ and an assumed model $H$. This can be defined mathematically as

\begin{equation}\label{eq:bayes_theorem}
    p(\vec{\theta}|d, H) = \frac{p(d|\vec{\theta}, H)p(\vec{\theta}|H)}{p(d|H)},
\end{equation}

where $p(\vec{\theta}|d, H)$ is the posterior, $(d|\vec{\theta}, H)$ is the likelihood, $p(\vec{\theta}|H)$ is the prior and $p(d|H)$ is the evidence. The prior describes our knowledge about the parameters prior to any observations, in the context of gravitational-wave inference this is determined based on our understanding on the underlying astrophysics. The likelihood is the probability of the data for a given set of parameters. Finally, the evidence is the fully marginalised likelihood and expresses the probability observing the data given the model, irrespective of the parameters. Whilst computing the posterior is trivial in lower dimensions, the gravitational-wave signals from compact binary coalescence are described by a minimum of 15 parameters and the resulting parameter space can be multi-modal and highly correlated \cite{Veitch:2015}. Typically, this requires applying stochastic sampling techniques such as \ac{MCMC} \cite{Brooks:2011handbook} and Nested Sampling \cite{Skilling:2006}. These techniques are computationally expensive and their cost is directly related to the cost of evaluating the likelihood and amount of data being analysed. There have been various efforts to reduce this cost by means such as reparameterisations \cite{Farr:2014system-frame}, reduced order methods \cite{Purrer:2014,Smith:2016}, parallel sampling methods \cite{Handley:2015polychord,Veitch:2021cpnest,Smith:2020pbilby} and problem-specific sampling algorithms \cite{Veitch:2015,Lange:2018RIFT,Biwer:2018PyCBCInf,Ashton:2019}.

Machine learning algorithms have successfully been applied to various aspects of gravitational-wave data analysis  \cite{Cuoco:2020:egwml} including data-quality improvement, waveform modelling, searches and, most relevant to this work, parameter estimation \cite{Gabbard:2019,Chua:2019,Green:2020a,Green:2020complete}. These algorithms, once trained, greatly reduce the cost of computing posterior distributions. However they are still in their infancy and there are various challenges that have yet to be solved, for example, they are constrained by the distribution of training data and cannot reliably be applied to data that is outside that distribution, e.g. with different detector noise curves or longer duration signals.

Normalising flows are a type of generative machine learning algorithm that have recently been applied to a wide variety of problems including generation, inference and representation learning \cite{Kobyzev:2019nf,Paramakarios:2019nfpmi}. The fundamental idea behind normalising flows is to model a complex probability distribution as the transformation of simple distribution. This transformation is constructed to be highly flexible but maintain an explicit mathematical description with a tractable Jacobian. This makes them well suited to applications in the physical sciences where the apparent opacity of other machine learning algorithms can hinder their widespread adoption.

In this paper we propose a modified nested sampling algorithm that incorporates a novel proposal method using normalising flows. During sampling the normalising flow is trained such that the resulting distribution can be re-sampled according to the prior to produce new live points within a given iso-likelihood contour which are independent of the current \textit{worst point}. This is akin to other nested sampling algorithms which sample from the constrained prior thus avoiding the need to evolve points with a random walk. These algorithms typically require fewer evaluations of the likelihood compared to those that use random walks which reduces the computational cost. However, the challenge when implementing these algorithms is determining the constrained prior which often requires defining multiple bounding distributions. Using a normalising flow allows us to use a single bounding distribution but comes at the cost of training during sampling. This additional cost is easily offset in problems where there is significant computational cost associated with computing the likelihood. As such we choose to apply our algorithm to gravitational-wave inference.

This paper is structured as follows: in \cref{sec:background} we outline the background theory for gravitational-wave inference, nested sampling and normalising flows. We then introduce our method in \cref{sec:method} and discuss related work in \cref{sec:related_work}. In \cref{sec:results} we present results for gravitational-wave parameter estimation, compare to a commonly used nested sampler, and discuss the implication of our results. Finally in \cref{sec:conclusion} we summarise our results and draw conclusions.

\section{Background}\label{sec:background}

\subsection{Gravitational-wave likelihood for compact binary coalescence}\label{sec:gw_likelihood}


The strain data $\vec{d}$ from gravitational-wave interferometers is modelled as a signal with additive noise in the time domain \cite{Veitch:2015}. The standard definition of the likelihood for compact binary coalescence then assumes that the noise is Gaussian and stationary over short observation times and that the power spectral density $S_n$ is independent of the model parameters. This allows us to define the likelihood as an independent product in the frequency domain
\begin{equation}\label{eq:gw_likelihood}
    p(\vec{d}|H, \vec{\theta}) \sim \exp \sum_{i} \left[ \frac{-2|\tilde{h}_{i}(\vec{\theta}) - \tilde{d}_{i}|^{2}}{TS_{n}(f_{i})} \right],
\end{equation}
where $T$ is the duration of the data $\vec{d}$ in seconds and $\tilde{h}_{i}(\vec{\theta})$ and $\tilde{d}_{i}$ are the waveform model and data in the frequency domain. There are various different waveform models $\vec{h}(\vec{\theta})$ which are characterised by sets of parameters. The exact number of parameters depends on the physics that the model describes, if both compact objects are assumed to be black holes, then typically 15 parameters are required \cite{Veitch:2015}. Different parameterisations have proven to improve sampling efficiency and convergence, most notably reparameterising the component masses in terms for chirp mass $\mathcal{M} = (m_1 m_2)^{3/5} (m_1 + m_2 )^{-1/5}$ and asymmetric mass ratio $q = m_{2} / m_{1}$ \cite{Veitch:2015} and using the \textit{system-frame} when describing the orientation of the binary \cite{Farr:2014system-frame}.

\subsection{Nested sampling}\label{sec:nested_sampling}

Nested sampling is a stochastic algorithm for Bayesian inference that was proposed by Skilling in \cite{Skilling:2006} and computes the Bayesian evidence $Z = p(\vec{d}|H) = \int p(\vec{d}|\vec{\theta}, H) \diff\vec{\theta}$ in \cref{eq:bayes_theorem}. In nested sampling the evidence integral is simplified by considering the total prior volume $X$ contained within a given iso-likelihood contour $L^{*} = p(\vec{d}|\vec{\theta}, H)$:

\begin{equation}\label{eq:ns_prior_volume}
    X(L^{*}) = \int_{p(\vec{d}|\vec{\theta}, H) > L^{*}} \diff \vec{\theta} p(\vec{\theta}|H).
\end{equation}

where $\diff \vec{\theta} p(\vec{\theta}|H) = dX$. The evidence integral can then be re-written using the likelihood $L$ by inverting \cref{eq:ns_prior_volume}

\begin{equation}\label{eq:ns_evidence}
    Z = \int_{0}^{1} L(X)\diff X.
\end{equation}

Since the integrand is positive and decreasing, the function is well behaved and the integral can therefore be approximated by considering an ordered sequence of decreasing of $M$ points in the prior volume $X_{i}$, evaluating the likelihood at each point $L_{i} = L(X_{i})$ and finally, for example, using the trapezoid rule

\begin{equation}\label{eq:ns_trapezoid}
    Z = \sum_{i}^{M} \frac{1}{2} (X_{i-1} - X_{i+1}L_{i}).
\end{equation}

The complete algorithm is detailed in \cite{Skilling:2006} but in short it requires first sampling a set of points, known as live points, from the prior distribution. The point with the lowest likelihood $L^{*}$ is then removed and replaced by another sample drawn from within the likelihood contour defined by the original point and according to the prior distribution. This replacement process is then continued until a stopping criteria is met \cite{Skilling:2006}. In gravitational-wave inference this is typically chosen to be when the fractional change in log evidence at a given iteration $\Delta \ln Z_{i}$ caused by replacing another point is less than $0.1$ \cite{Veitch:2015}, where $\Delta \ln Z_{i} \equiv \ln \left(Z_{i} + \Delta Z_{i}\right) - \ln Z_{i}$ and $\Delta Z_{i} \approx L^{*} X_{i}$. Once the algorithm has terminated the final evidence is computed including the final set of live points and posterior samples can be produced from the set of discarded and final live points \cite{Skilling:2006}. For a recent review of nested sampling see \cite{Buchner:2021}.

The challenge when implementing this algorithm is efficiently proposing the replacement samples within the current iso-likelihood contour since these samples must be \ac{iid} according to the prior. One commonly used method draws new points based on the location of existing points which are then evolved using a random walk to ensure they are \ac{iid} \cite{Veitch:2015,Veitch:2021cpnest}. The efficiency of this method depends on the length of the \ac{MCMC} chain required since each step in the chain requires evaluating the likelihood, for \ac{GW} inference proposing a single new sample often requires of order $10^{3}$ likelihood evaluations. Other approaches avoid using a random walk by sampling from the constrained prior at a given iteration. This requires determining the distribution of the current live points and constructing one or more bounding distributions, for example, using multidimensional ellipsoids as described in \cite{Feroz:2019multinest}. In this case the challenge is efficiently constructing the bounding distributions and sampling them without over-constraining the prior and consequently under-sampling a region in parameter space. What is more, these two approaches are sometimes used in conjunction \cite{Speagle:2020} to tackle particularly challenging problems, such as gravitational-wave inference \cite{Romero-Shaw:2020}, where the individual methods can be inefficient. In \cref{sec:method} we propose a method for directly sampling from a constrained prior using normalising flows to learn the iso-likelihood contours.

\subsection{Normalising flows}

Normalising flows \cite{Rezende:2015} are a type of generative machine learning algorithm that map samples $x$ in the physical space\footnote{We use the term physical space to distinguish it from the strain data however in the literature $\physical$ often referred to as the data space.} $\physical$ to samples $z$ in a latent space $\latent$ such that the samples $z$ are distributed according to a prior probability distribution $p_{\latent}$, known as the \textit{latent prior}. If we assume this mapping to be bijective and denoted $f: \latent \to \physical$, then using the change of variable formula we can define a probability distribution in the physical space $p_{\physical}$

\begin{equation}\label{eq:change_of_variable}
    p_{\physical}(x) = p_{\latent}(f(x)) \left| \textrm{det} \left( \frac{\partial f(x)}{\partial x^{T}}\right)\right|,
\end{equation}

where $\partial f(x) / \partial x^{T}$ is the Jacobian of $f$ at $x$. Samples can then be drawn from $p_{\physical}$ by sampling the latent space $z \sim p_{\latent}$ and applying the inverse mapping $f^{-1}$. This requires the bijective mapping $f$ to have a tractable Jacobian, which for functions with multidimensional domains and codomains is often computationally expensive. These mappings are therefore carefully constructed to meet these criteria and usually parameterised by a neural network which allows for the mapping to be learned.

Normalising flows generally fall in two categories depending on how the mappings are defined: \textit{autoregressive flows} \cite{Papamakarios:2017made,Huang:2018naf} and flows based on \textit{coupling transforms} \cite{Dinh:2016:rnvp,Kingma:2018glow,Durkan:2019spl}. Each have distinct advantages and disadvantages, most notably autoregressive flows are often more flexible but have a greater computational cost and are not always analytically invertible whereas coupling based flows tend to be less flexible but computationally cheaper to evaluate and analytically invertible. In this work we use the latter since we aim to minimise computational cost and later rely on the invertablility of normalising flows.

\begin{figure*}
    \centering
    \includegraphics[width=\onehalffigwidth]{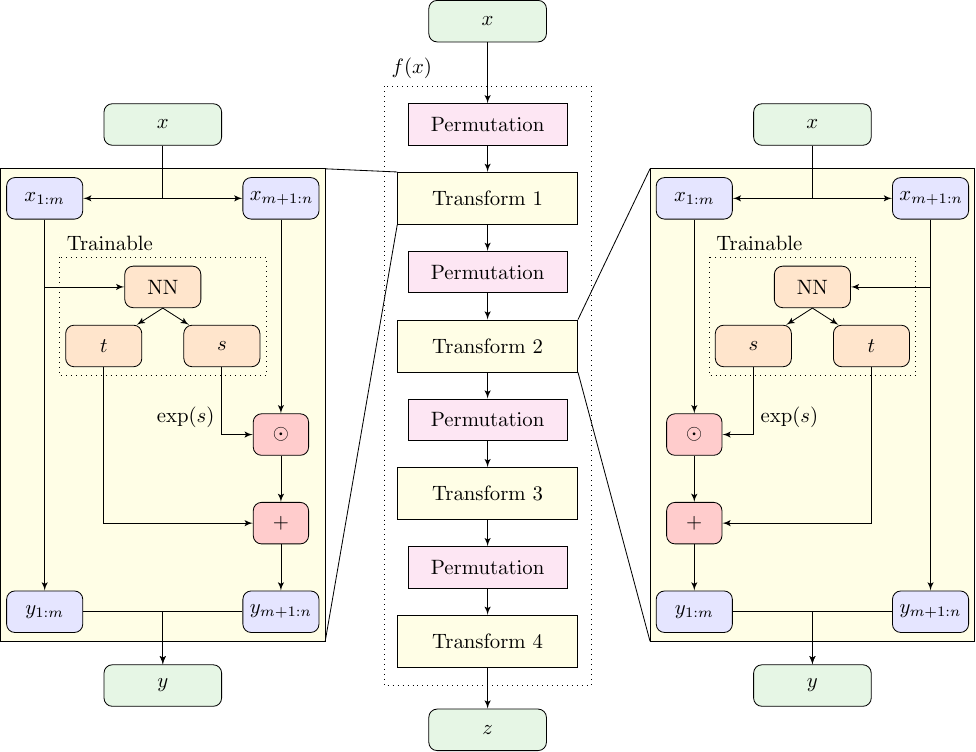}
    \caption{Diagram of a normalising flow $f(x)$ composed of four coupling transforms which maps an \protect\ndimensional{n} input vector x to an \protect\ndimensional{n} latent vector z. Each transform splits $x$ in two $[x_{1:m}, x_{m+1:n}]$ and updates one part conditioned on the other. In the first and third transforms $x_{1:m}$ is used as the input to a neural network (NN) which then produces the scale $s$ and translation $t$ vectors of length $m$. The element-wise product ($\odot$) is then computed between $x_{1:m}$ and $\exp(s)$ followed by the sum of the output and $t$. This is shown in the left transform. In the second and fourth transforms $x_{1:m}$ is updated conditioned on $x_{m+1:n}$ as shown in the right transform.}
    \label{fig:flow_diagram}
\end{figure*}

Flows based on coupling transforms map an \ndimensional{n} input vector $x$ to an \ndimensional{n} output vector $y$ by splitting the input vector in two parts $x = [x_{1:m}, x_{m+1:n}]$ where $m < n$ and then transforming one part conditioned on the other unchanged part. The choice of splitting depends on the specific coupling transform though typically an alternating pattern is used. For example the coupling transform proposed in \cite{Dinh:2016:rnvp}:

\begin{subequations}\label{eq:coupling_trasnform}
    \begin{align}
        y_{1:m} & = x_{1:m}, \\
        y_{m+1:n} & = x_{m+1:n} \odot \exp \left[ s(x_{1:m})\right] + t(x_{1:m}),  
    \end{align}
\end{subequations}

where $s(x_{1:m})$ and $t(x_{1:m})$ are the output of a neural network and $\odot$ denotes the element-wise product. We show a schematic of \cref{eq:coupling_trasnform} in \cref{fig:flow_diagram}. Since $y_{1:m} = x_{1:m}$ the Jacobian matrix of these transforms is lower triangular and  the Jacobian determinant is simply the product of the diagonal entries. This transform can easily be inverted by inverting \cref{eq:coupling_trasnform} and using $y_{1:m}$ as the input to $s$ and $t$:

\begin{subequations}\label{eq:inverse_coupling_trasnform}
    \begin{align}
        x_{1:m} & = y_{1:n}, \\
        x_{m+1:n} & = \left[y_{m+1:n} - t(y_{m:1})\right] \odot \exp \left[-s(y_{1:m})\right].  
    \end{align}
\end{subequations}

As with \cref{eq:coupling_trasnform} the Jacobian determinant is trivial to compute. Coupling transforms can also be stacked by alternating which part of $x$ is updated, $x_{1:m}$ or $x_{m+1:n}$, to produce more flexible transforms. It is also common practice to include permutations between coupling transforms, these permute the inputs to the transforms making the mapping more expressive \cite{Dinh:2016:rnvp}. More recently this has been generalised to \textit{linear transforms} in which the permutation is learnt during training \cite{Kingma:2018glow}. In \cref{fig:flow_diagram} we show a schematic of how coupling transforms can be stacked with permutations. For a more detailed description of normalising flows, the different types and their application we point the reader to \cite{Kobyzev:2019nf,Paramakarios:2019nfpmi}.

\section{Method}\label{sec:method}

We present a novel method for sampling within a given iso-likelihood using a normalising flow. The normalising flow learns the distribution of a set of live points and is constructed such that the learnt distribution can be sampled from analytically. We introduce additional steps to ensure that the samples are distributed according to the \textit{sampling prior}\footnote{We use sampling prior to denote the prior used for nested sampling and to distinguish it from the latent prior use in normalising flows.} and bounded by the iso-likelihood contour. This eliminates the need to evolve new samples and allows us to efficiently draw new samples within a complex iso-likelihood contour. A more efficient proposal equates to few rejected points and, since the likelihood must be computed before a point can be rejected, this is also equivalent to a reduction in the number of likelihood evaluations. We first describe our method in the isolated case of a single of set of live points at a given iteration and then present a nested sampling algorithm which incorporates it.

\begin{figure}
    \centering
    \includegraphics[width=\figwidth]{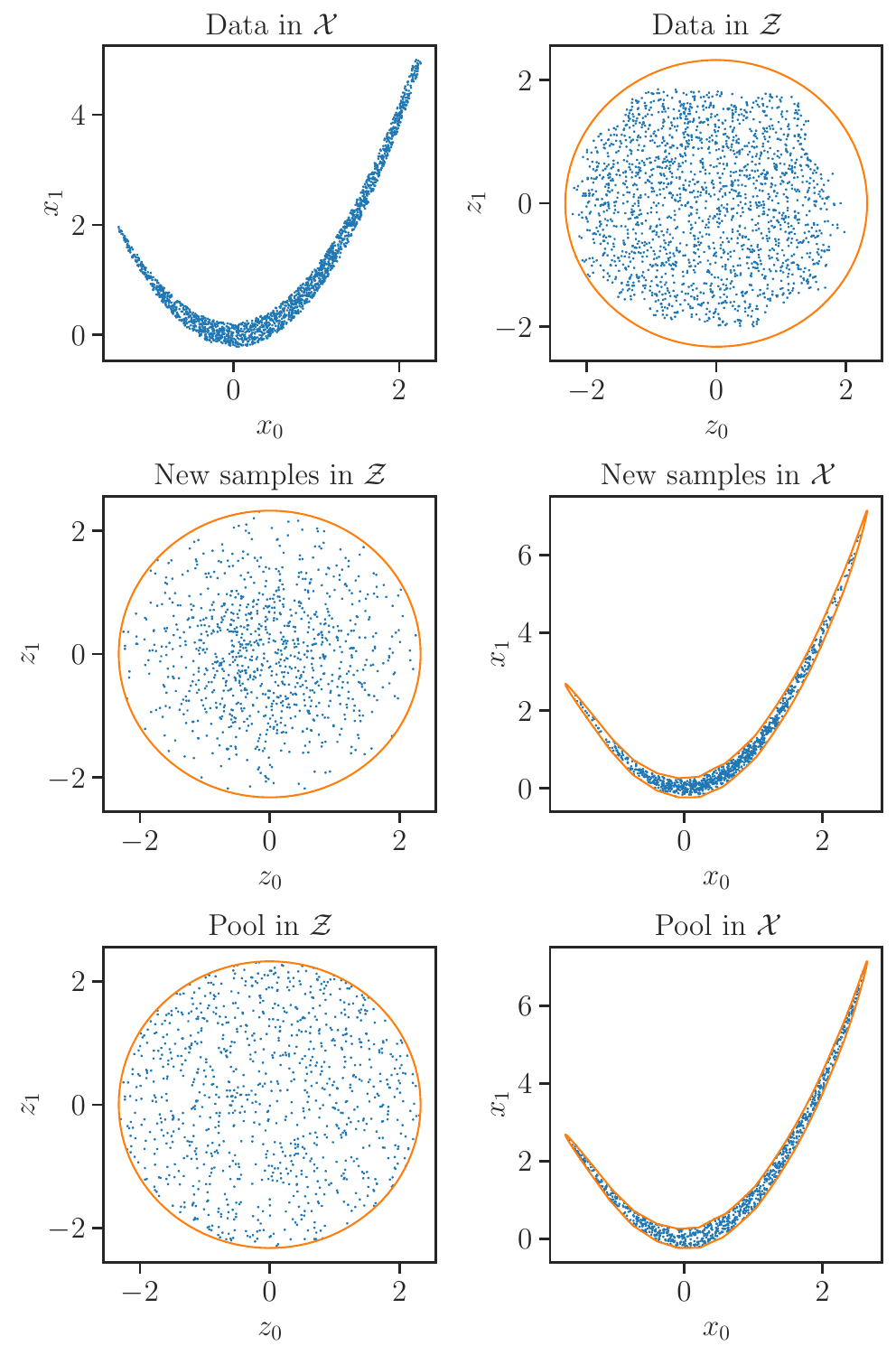}
    \caption{Example of how a normalising flow trained on a set of live points can produce samples within current iso-likelihood contour for simple two-dimensional parameter space. \textbf{Top:} example of training samples in the physical space $\physical$ and learned mapping to the latent space $\latent$ with the iso-likelihood contour for the current \textit{worst point} shown in orange. \textbf{Middle:} samples drawn from a truncated Guassian within the iso-likelihood contour in $\latent$ and mapped to $\physical$ using the inverse mapping. \textbf{Bottom:} pool of accepted samples after applying rejection sampling until 1000 points are obtained shown in both $\latent$ and $\physical$.}
    \label{fig:learning_contours}
\end{figure}

\subsection{Sampling within a iso-likelihood contour}\label{sec:sampling}

At any given point in the nested sampling algorithm there is a current set of live points which by definition is contained within the iso-likelihood contour defined by the \textit{worst point} with likelihood $L^{*}$. We describe the implementation in our algorithm in terms of four steps:

\paragraph{How to define an iso-likelihood contour using a normalising flow.} 

If we treat the sampling space as the physical space $\physical$ and the live points as the data $x$ we can then train a normalising flow to approximate the distribution of the live points to within some error and use it to draw new samples in $\physical$. This requires sampling from the latent prior $p_z$, which we choose to be an \ndimensional{n} Gaussian, and then applying the inverse mapping learned by flow $f^{-1}$. However, since our choice of latent prior $p_{\latent}$ has an infinite domain the resulting distribution of samples $p_{\physical}$ will not be bounded by the iso-likelihood contour or distributed according to the sampling prior. We therefore examine the notion of an iso-likelihood contour in the context of the normalising flow with an \ndimensional{n} Gaussian latent prior.

\paragraph{How to determine the contour given the current set of live points.} 

Once trained, the normalising flow can be used to map the current \textit{worst point} in the physical space $x^{*}$ to the latent space $\latent$ as $z^{*}$. This point has a likelihood in the latent space $L_{\latent}^{*}$ given by $p_\latent(z^*)$ and, since $p_\latent$ is an \ndimensional{n} Gaussian, points of equal likelihood lie on the \nsphere{(n-1)} with radius $r^{*}$ given by $z^{*}$. If we assume a perfect mapping, then this iso-likelihood contour in the latent space can be mapped to an iso-likelihood contour in the physical space. We can therefore sample within the contour in latent space and use the inverse mapping to produce samples within the  contour in the data space.

\paragraph{How to sample within the contour.}

We use two approaches for drawing $K$ new samples $\vec{z}_{i}$ in the latent space given a radius $r^{*}$, these produce normally and uniformly distributed samples respectively. Both start by drawing $K$ samples on the \nsphere{(n-1)} using the algorithm proposed in \cite{Muller:1959ANO, Marsaglia:1972} where \ndimensional{K} vectors are drawn from an \ndimensional{n} unit Gaussian and then normalised using the Euclidean norm. These samples $\vec{y}_{i}$ can then be rescaled to obtain samples within in the \nball{n}

\begin{equation}\label{eq:latent_sampling}
    \vec{z}_{i} = \rho_{i} \frac{\vec{y}_{i}}{||\vec{y}_{i}||_{2}},
\end{equation}

where the choice of distribution for $\rho_{i}$ determines how the resulting samples $\vec{z}_{i}$ are distributed in $\latent$. For uniformly distributed samples $\rho_{i} = u^{1/n}$ where $u\sim \mathcal{U}(0, r^{*})$ and for normally distributed samples $\rho_{i} \sim \chi(n)$ where $\chi(n)$ is a chi-distribution with $n$ degrees of freedom truncated at $r$. The inverse mapping of the normalising flow can then be applied to $\vec{z}_{i}$ to obtain samples in the physical space. We consider two approaches because sampling from a truncated Gaussian in high-dimensional space can become inefficient for large values of $r^{*}$.

\paragraph{How to ensure new samples are drawn according to the prior.}

The samples obtained in the previous step must be re-sampled such that they are distributed according to the sampling prior. We use rejection sampling and compute weights $\alpha_{i}$ for each sample

\begin{equation}\label{eq:rejection_sampling}
    \alpha_{i} = \frac{p(\vec{x}_{i})}{q(\vec{x}_{i})},
\end{equation}

where $x_{i} = f^{-1}(z)$, $p(x)$ is the sampling prior and $q(x)$ the proposal probability which is computed using the inverse of \cref{eq:change_of_variable}

\begin{equation}\label{eq:proposal_prob}
    q(x) = q(f^{-1}(z)) = p_{\latent}(z)\left| \textrm{det} \left( \frac{\partial f^{-1}(z)}{\partial x^{T}}\right)\right|^{-1}.
\end{equation}

The choice of latent prior $p_{\latent}$ will depend on which method was used to draw the samples in the latent space. The weights \cref{eq:rejection_sampling} are then rescaled such that their maximum value is one. We then draw $N$ samples $u \sim \mathcal{U}[0, 1]$ and accept samples for which $\alpha_{i}/u_{i} > 1$. In \cref{fig:learning_contours} we show an example of this process for a simple two-dimensional case.

\subsection{Algorithm details}\label{sec:algorithm}

We identify three key stages of the sampling approach detailed in \cref{sec:sampling}  that we then incorporate into the nested sampling algorithm:

\begin{itemize}
    \item {\textbf{Training:} a normalising flow is trained on the current set of $K$ live points by minimising a Monte Carlo approximation of the of the KL divergence between the target distribution and $p_{\physical}(x)$ as defined in \cref{eq:change_of_variable}. The explicit loss function is derived in \cref{app:loss}.}
    \item {\textbf{Population:} once the normalising flow is trained, samples are drawn within the $n$-ball of radius $r$ defined by the \textit{worst point}. These samples are then mapped to the data space $\physical$, re-sampled according to the prior and finally stored in the pool of new samples.}
    \item {\textbf{Proposal}: once the pool of new samples has been populated, new live points are drawn at random from the pool and then removed until the pool is empty or the normalising flow is retrained.}
\end{itemize}

The standard nested sampling algorithm is modified to include these stages, we call the algorithm \nessai. 

The start of the algorithm remains unchanged: K live points are drawn from the prior distribution and their log-likelihoods computed. We then start the iterative process of determining the \textit{worst} live point with log-likelihood $L^{*}$ and drawing a replacement live point that lies within the iso-likelihood contour. In our modified algorithm we use standard rejection sampling from the prior for the first $M$ points (typically $2K$) or until it becomes inefficient. The normalising flow is then trained on the current $K$ live points allowing us to map the \textit{worst point} to the latent space to obtain the \textit{worst latent point} $z_{w}$ and the radius of the corresponding \nball{n} $r$. This radius is then used for the population stage when drawing samples in the latent space. Once populated, a replacement point is drawn from the pool, its log-likelihood is computed and if it is greater than $L^{*}$, the point is accepted; if not, more points are drawn until one is accepted. The proposal stage is then repeated for subsequent worst live points until one of four criteria is met:

\begin{itemize}
    \item {\textbf{the proposal pool is depleted:} the normalising flow is retrained using the current live points and  \textit{worst} live point is used to compute a new radius and the population stage is repeated,}
    \item {\textbf{the acceptance rate falls below a user-defined criteria:} the current proposal pool is discarded and the normalising flow is retrained. This threshold is defined by the user,}
    \item {\textbf{the criteria for retraining the normalising flow is met:} the normalising flow is retrained with the current $K$ live points, this happens by default every $K$ live points,}
    \item {\textbf{the nested sampling convergence criterion is met:} the algorithm terminates.}
\end{itemize}

Since this algorithm relies on the normalising flows' ability to approximate the distribution of live points at various stages throughout the sampling, we include a series of reparameterisations of $\physical$ to reduce the complexity of the data space and removing certain features. We denote this reparameterised space $\physical'$ and include the Jacobian for each reparameterisation in \cref{eq:proposal_prob}. These reparamerisations are:

\begin{itemize}
    \item {\textbf{Rescaling:} we add the option to rescale the input data according to either the \textit{sampling priors} or the current minimum and maximum values such that all of the parameters in $\physical'$ are defined over the same domain. As as default we use $[-1, 1]^{n}$.}
    \item {\textbf{Boundary inversion:} we observe that asymmetric distributions with high density regions near the prior bounds are often under-sampled. To mitigate this effect we add the option to \textit{mirror} the live points around such bounds and train on the resulting symmetric distribution. Further details are provided in \cref{app:inversion}.}
\end{itemize}

We also introduce additional settings which help with convergence and sampling efficiency, some of these and discussed in \cref{sec:tuning} and a comprehensive list can be found in the online documentation for our sampler \cite{nessai-docs}.

\subsection{Gravitational-wave reparameterisations}\label{sec:gw_reparam}

The gravitational-wave parameter space is typically \ndimensional{15} and contains various degeneracies between parameters such as the masses, inclination and luminosity distance which can make sampling inefficient. Previous work has shown that certain reparameterisations can improve sampling efficiency \cite{Veitch:2015}.  We use two of these: chirp mass $\mathcal{M}$ and asymmetric mass ratio $q$ replace the component masses and we use the \textit{system-frame} parameterisation in place of the \textit{radiation-frame} to describe the orientation of the binary \cite{Farr:2014system-frame}.

More than half of the parameters to sample are angles and we note that the periodicity of these angles is not encoded in the mapping learned by the normalising flow since the latent space $\latent$ is continuous and unbounded. We therefore include a further reparameterisation specifically for the angular parameters $\theta_i$. We assume that each angle has a corresponding radial component $\rho_{\theta_{i}}$ and together they describe a position in a two-dimensional plane. We can therefore use standard transformations to express this position in Cartesian coordinates $(x_{\theta_{i}}, y_{\theta_{i}})$:

\begin{equation}\label{eq:cartesian}
    \begin{split}
    x_{\theta_{i}} = \rho_{\theta_{i}} \cos{\theta_{i}}, \\
    y_{\theta_{i}} = \rho_{\theta_{i}} \sin{\theta_{i}}.
    \end{split}
\end{equation}

If we choose the distribution of radial components such that $\rho_{\theta_{i}} \in [0, \infty)$ then $x_{\theta_{i}}, y_{\theta_{i}} \in (-\infty, \infty)$. Since we are using a Gaussian latent prior $p_\latent$, we sample $\rho_{\theta_i}$ from a chi-distribution with two degrees of freedom such that, if the angle is uniformly distributed on $[0, 2 \pi]$, the resulting distribution of $(x_{\theta_{i}}, y_{\theta_{i}})$ is Gaussian. We use this treatment for the phase, inclination, polarisation and all four spin angles, for polarisation we rescale the angles to $[0, 2\pi]$ before applying the transformation to Cartesian coordinates. This reparameterisation also naturally includes periodic boundary conditions for the angles with uniform priors.

The sky location is described by a further two angles, right ascension $\alpha$ and declination $\delta$. For these angles we extend the previous treatment from two-dimensional to three-dimensional Cartesian coordinates $(x, y, z)$ and draw the radial component $\rho$ from a chi-distribution with three degrees of freedom. For the standard priors, $p(\alpha) \sim \mathcal{U}[0, 2 \pi]$ and $p(\delta) \sim \cos{\delta}$, the resulting distribution of $(x, y, z)$ is again Gaussian. 


The spin magnitudes $\chi_1$ and $\chi_2$ also require a specific treatment. They are typically defined on $[0, 0.99]$ with uniform priors and, importantly, the posterior distributions are often broad and span the entire prior range. We consider applying the boundary inversion to both bounds but in practice find this ineffective. We instead opt to map $\chi_i$ into a two-dimensional plane with positions described using Cartesian coordinates $x_{\chi_i}$ and $y_{\chi_i}$. We achieve this by first defining a rescaled magnitude $\hat{\chi}_{i} \in [0, 1]$ which is obtained using the corresponding priors. Then, we consider the angle defined by $\hat{\chi_{i}}\pi$ and, again, introduce a radial component $\rho_{\chi_{i}} \sim \chi(2)$. The corresponding Cartesian coordinates $(x_{\chi_i}, y_{\chi_i})$ are defined on $[0,\infty)$ and $(-\infty, \infty)$ respectively. However we know that the coupling transforms we have chosen to use are better suited to unbounded domains. To avoid this, we introduce a random variable $k$ which is drawn from a Rademacher distribution and include it in the Cartesian coordinate transform 

\begin{equation}\label{eq:spin_reparam}
    \begin{split}
        x_{\chi_{i}} & = \rho_{\chi_{i}} \cos{\hat{\chi_{i}}\pi}, \\
        y_{\chi_{i}} & = k\rho_{\chi_{i}} \sin{\hat{\chi_{i}}\pi}.
    \end{split}
\end{equation}

As a result of the including $k$, $(x_{\chi_i}, y_{\chi_i}) \in (-\infty, \infty)$ and the hard boundary at $x_{\chi_{i}}=0$ has been avoided.

We choose to reparameterise the luminosity distance $d_{\text{L}}$ such that the prior for the resulting parameter $d_{\text{U}}$ is uniform. The exact reparameterisation therefore depends on the prior used for $d_{\text{L}}$. In this work we choose to use a prior on $d_{\text{L}}$ that is uniform in co-moving volume so we first convert the luminosity distance to a co-moving distance $d_{\text{C}}$ and then the uniform parameter is simply $d_{\text{U}} = d_{\text{C}}^{3}$. Similar reparametersitions can be determined for other commonly used distances priors such as a power law. Additionally, we allow boundary inversion as described in \cref{sec:algorithm} but limit it to only the upper bound since in practice the luminosity distance posterior will not rail against the lower bound.

The remaining parameters are the chirp mass $\mathcal{M}$ and mass ratio $q$ for which we use the reparameterisations from $\physical$ to $\physical'$ mentioned in \cref{sec:algorithm}, allowing boundary inversion for $q$.


\subsection{Implementation}

We use the implementation of normalising flows in \pytorch \cite{Paszke:2019:pt} available in \nflows \cite{nflows} which allows for a wide variety of normalising flows to be used. However we choose to use coupling transforms \cite{Dinh:2016:rnvp} because of the tractable Jacobian and ease of computing the inverse mapping.  As suggested in \cite{Dinh:2014nice,Kingma:2018glow}, we include invertible linear transforms that randomly permute the parameters before each coupling transform allowing all of the parameters to interact with each other. We also include batch normalisation \cite{Ioffe:2015:bn} after each coupling transform as described in \cite{Dinh:2016:rnvp}. We use a residual neural network \cite{He:2015:drl, He:2016:imdrn} for computing the parameters for each transform. We train the normalising flows with the Adam optimiser \cite{Kingma:2014}. In \cref{app:sampling-settings} we detail the specific parameters used for the results presented in \cref{sec:results}.

Our sampler, \nessai (Nested Sampling with Artificial Intelligence), is available as an open source package \cite{nessai-git} and documentation is also available online \cite{nessai-docs}.

\section{Related Work} \label{sec:related_work}

Different frameworks and samplers have been developed for gravitational-wave inference. \lalinference \cite{Veitch:2015} implements nested sampling and \ac{MCMC} with specific proposal methods for the gravitational-wave parameter space and has been used extensively for analyses of the first gravitational wave detections \cite{Abbott:2016blz} and GWTC-1 \cite{LIGOScientific:2018mvr}. More recently, the \python package \bilby \cite{Ashton:2019} has been developed to use off-the-shelf samplers, such as \dynesty \cite{Speagle:2020}, and been shown to achieve comparable results to \lalinference on GWTC-1 \cite{Romero-Shaw:2020}.

Machine learning has previously been incorporated into stochastic sampling algorithms; in \cite{Graff:2013bambi} the likelihood function is approximated with a neural network, and in \cite{Levy:2017ghmc} neural networks are used to generalise Hamiltonian Monte Carlo. More closely related to our work, normalising flows have been used to improve the efficiency of \ac{MCMC} methods by reparameterising the sampling space \cite{Hoffman:2019} and a similar approach has also been extended to \ac{MCMC} sampling in nested sampling in \cite{Moss:2019}.

Recent work has shown that likelihood-free inference using conditional variational autoencoders \cite{Gabbard:2019, Chua:2019} and normalising flows \cite{Green:2020a, Green:2020complete} can produce posterior distributions for compact binary coalescence from binary black holes. These approaches promise to drastically reduce the cost of producing posterior samples when compared to traditional stochastic sampling methods. However, they require large amounts of training data and they currently lack the flexibility to deal with, for example, different PSDs, high sampling frequencies and long duration signals.

\begin{figure*}
    \subfigure[With phase marginalisation]{\includegraphics[width=\figwidth]{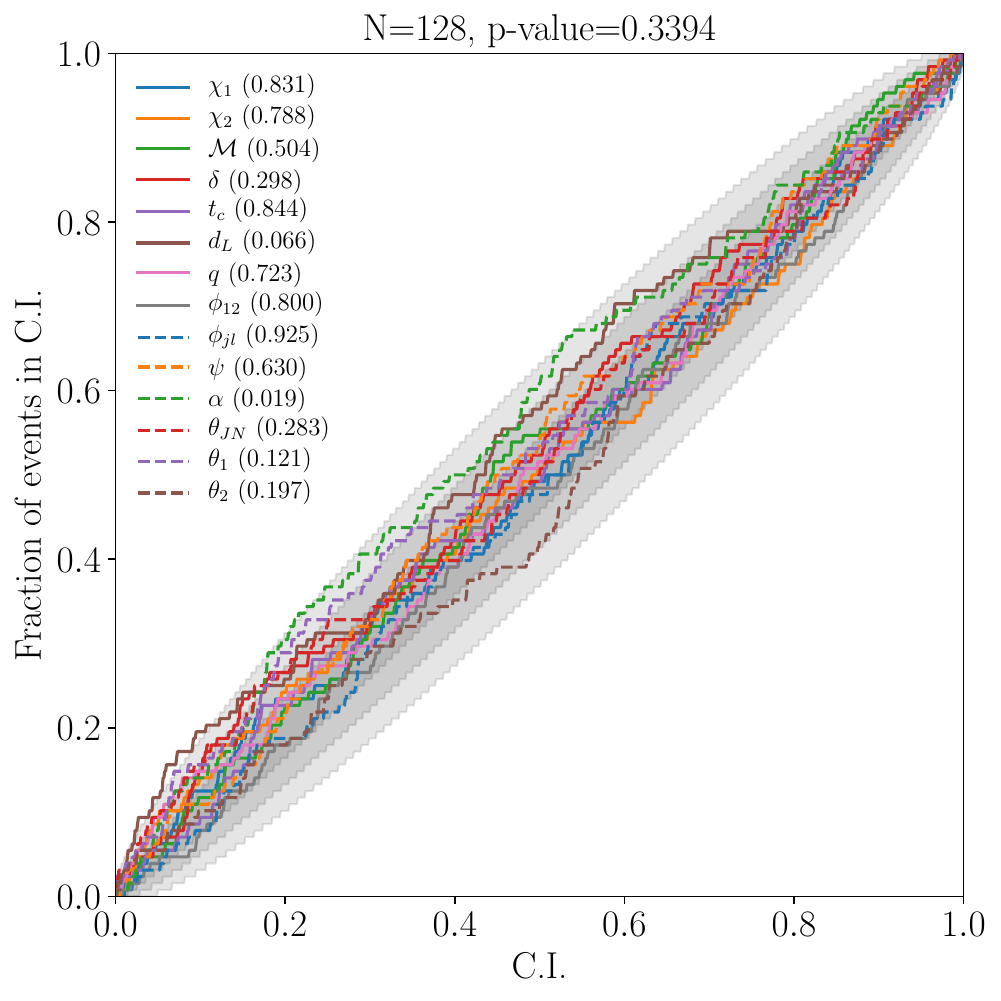}}
    \subfigure[With phase and distance marginalisation]{\includegraphics[width=\figwidth]{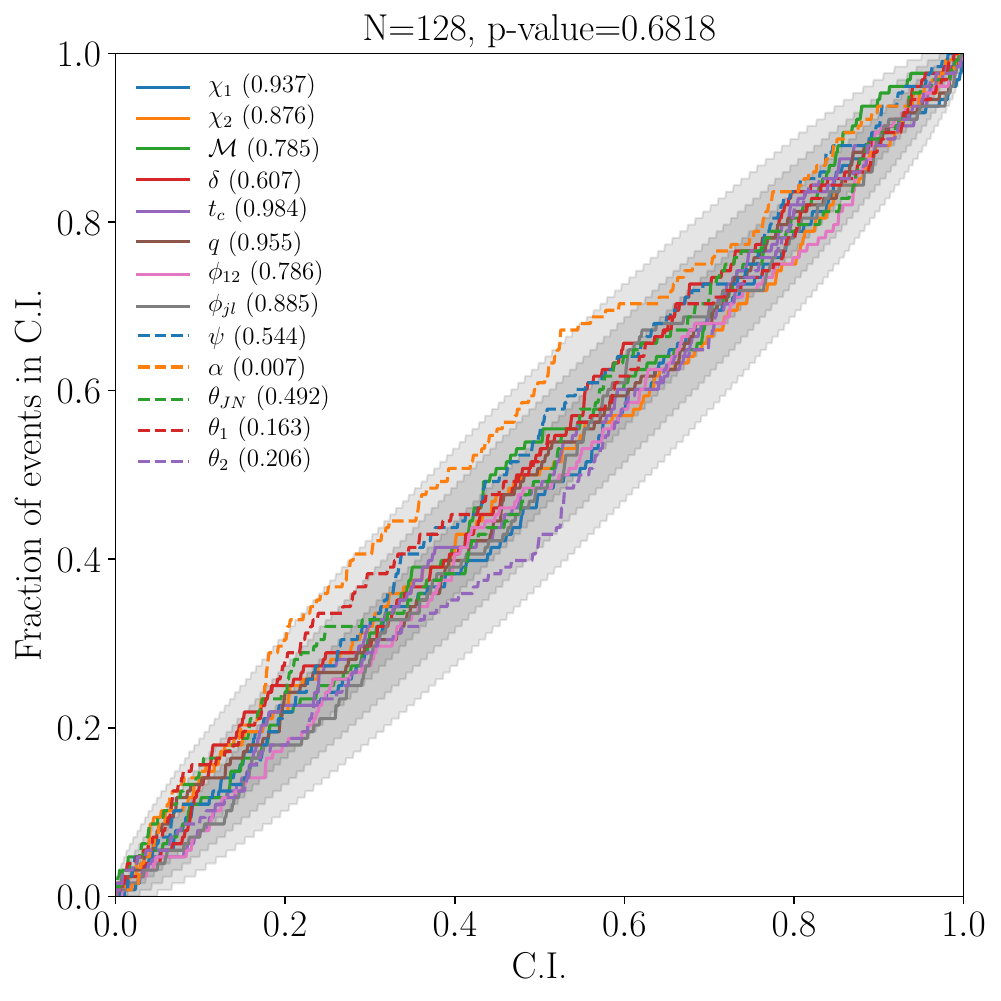}}
    \caption{Probability-probability (P-P) plot showing the confidence interval versus the fraction of the events within that confidence interval for the posterior distributions obtained using our analysis \nessai for 128 simulated compact binary coalescence signals produced with \bilby and \bilbypipe. The 1-, 2- and 3-$\sigma$ confidence intervals are indicated by the shaded regions and $p$-values are shown for each of the parameters and the combined $p$-value is also shown.}
    \label{fig:pp_plot}
\end{figure*}

\section{Results}\label{sec:results}

We chose to evaluate the performance of our sampling algorithm, \nessai, with simulated gravitational-wave signals from \ac{CBC}. The parameter is multi-dimensional and various parameters are correlated which can prove challenging when sampling. The likelihood, as defined in \cref{sec:gw_likelihood}, is also typically computationally costly to evaluate, making it well suited to our sampler, however, this depends on the waveform approximant used and length of the observation. We first use probability-probability plots to check the consistency of our sampler and then compare our results to those obtained using \dynesty. We then highlight how the likelihood computation can be parallelised in \nessai before finally discussing various diagnostics that can be used to identify problems during sampling and tune the sampler settings.

\begin{figure}
    \centering
    \includegraphics[width=\figwidth]{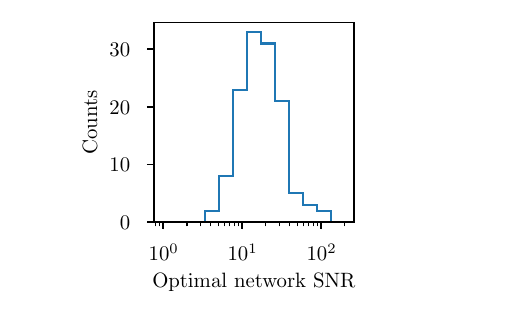}
    \caption{Distribution of the optimal network \ac{SNR} for the 128 simulated gravitational-wave injections in simulated Gaussian noise. The priors on chirp mass and luminosity distance \cref{app:priors} were chosen such that the ringdown frequency \cite{Abbott:2016bqf} does not exceed the Nyquist frequency and that the majority of signals have detectable optimal network \acp{SNR}}
    \label{fig:snr_dist}
\end{figure}

We use \bilbypipe and \bilby \cite{Ashton:2019} to simulate 128 injections using \imrphenomp \cite{Schmidt:2012,Khan:2019} sampled at 2048 Hz with 4-second observing time in a three detector network with AdLIGO Hanford, AdLIGO Livingston and AdVirgo at design sensitivity \cite{Aasi:2013wya}. We set the minimum frequency to 20 Hz and Gaussian noise is added to the injections using the PSDs for each detector. We choose uniform priors on chirp mass $\mathcal{M} \sim \mathcal{U} \unit[[25, 35]]{\text{M}_{\odot}}$ and asymmetric mass ratio  $q \sim \mathcal{U}[0.125, 1.0]$, a prior on luminosity distance that is uniform in co-moving volume on $\unit[[100, 2000]]\,{\text{Mpc}}$, a uniform prior for the reference time at the geocentre with width 0.2 and the remaining priors are set to the defaults for precessing binary black holes in \bilby \cite{Romero-Shaw:2020}, see \cref{app:priors} for a complete list. The specific priors on chirp mass and luminosity are chosen such that ringdown frequency \cite{Abbott:2016bqf} does not exceed the Nyquist frequency and the majority of signals have detectable optimal network \acp{SNR}, the distribution of \acp{SNR} is shown in \cref{fig:snr_dist}.

We analyse the injections with our sampling algorithm, \nessai, outlined in \cref{sec:algorithm} and include the specific reparameterisations for gravitational wave analyses described in \cref{sec:gw_reparam}. We choose to analyse each injection twice: once with just phase marginalisation and once with both phase and distance marginalisation. Further details of the exact settings used for \nessai are provided in \cref{app:sampling-settings}.

\subsection{Result validation}\label{sec:validaton}

Probability-probability (P-P) plots are a standard method of verifying the performance of sampling algorithms \cite{Cook:2006pp,Talts:2018pp}. They test whether the correct proportion of injected values are recovered at a given confidence interval for a specific prior distribution. These tests are particularly useful when using a Gaussian likelihood, such as \cref{eq:gw_likelihood}, since the fraction of events within a given confidence interval should be uniformly distributed and we can therefore compute $p$-values for each parameter and a combined $p$-value for all of the parameters. We produce P-P plots for both of our analyses using \bilby and present the results in \cref{fig:pp_plot}. For an idealised sampler for the $p\%$ confidence interval, $p\%$ of the events should be recovered, this would correspond to a diagonal line. In practice we expect to see deviation from the diagonal, as such the 1-, 2- and 3-$\sigma$ confidence intervals are also shown in \cref{fig:pp_plot}. These results show that \nessai consistently recovers for the posteriors for the 128 injections but also indicate that luminosity distance is consistently harder to sample. The combined $p$-values of \nessaipvalue and \nessaipvaluemarg for our analyses without and with distance marginalisation serve as further verification.

\subsection{Comparison to \dynesty}

 To further validate our results we compare them to those obtained with \dynesty \cite{Speagle:2020}, another nested sampling algorithm commonly used in gravitational-wave inference \cite{Ashton:2019,Romero-Shaw:2020}. We use the configuration described in \cite{Romero-Shaw:2020} but increase the number of live points to 2000 and run on a single thread to ensure as direct of a comparison with \nessai as possible. With these settings \dynesty passes the P-P test (see \cref{app:dynesty}) but we note that these settings are the minimum required to produce reliable results and in practice more conservative settings are often used. Additionally, several injections required a second analysis with a different sampling-seed in order reach convergence. The results obtained with \dynesty allow us to verify the log-evidences returned by \nessai since these cannot be computed analytically and provides a point of reference when considering the number of likelihood evaluations and total computational time.

\begin{figure}
    \centering
    \includegraphics[width=\figwidth]{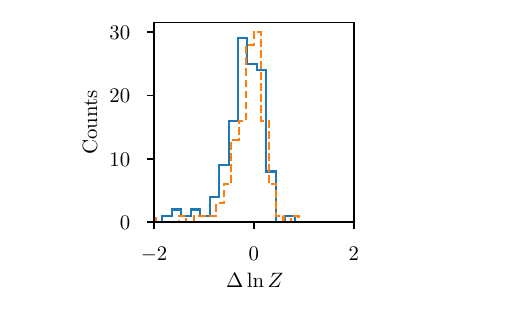}
    \caption{Difference between the log evidences $\Delta \ln Z$ obtained using \dynesty and \nessai for all 128 injections with distance marginalisation (dashed line) and without distance marginalisation (solid line).}
    \label{fig:log_evidence}
\end{figure}

In \cref{fig:log_evidence} we compare the log-evidences returned by \dynesty and \nessai. If \nessai was consistently over or under-estimating the log-evidence when compared to \dynesty, this would indicate a potential problem during sampling, such as over- or under-constraining, which would lead to biased results. The results in \cref{fig:log_evidence} show no such bias. However, since sampling is a stochastic process there is an error associated with the computed log-evidence. The theoretical error can be approximated using the information $H$ and the number of live points $K$, $\delta \log Z \approx \sqrt{H / K}$. To quantify this error we repeat the analysis on a single injection with 50 different sampling seeds and compute an approximate error $\delta \log Z \approx \logzerror$. In practice we observe a wider spread of log-evidences of $\logzsigma$, this is consistent with previous analyses which determined that there are additional sources of uncertainty \cite{Veitch:2010ns}.


\begin{figure}
    \centering
    \includegraphics[width=\figwidth]{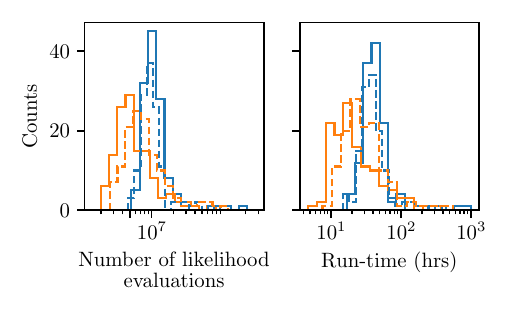}
    \caption{Distribution of the total number of likelihood evaluations required to reach convergence for and total run time for  \dynesty (blue) and \nessai (orange) when applied 128 simulated signals from compact binary coalescence with the priors and sampler settings described in \cref{sec:results,app:sampling-settings}. The results distance marginalisation disable are shown with solid lines and those with distance marginalisation enabled are shown with dashed lines.}
    \label{fig:comparison}
    
\end{figure}

\nessai is designed with the aim of improving the efficiency of drawing replacement live points at the cost of repeatedly training a normalising flow and populating a pool of live points. An improvement in the efficiency translates to a reduction in the total number of likelihood evaluations since the likelihood must be computed for each rejected point. We therefore compare the total number of likelihood evaluations required to reach convergence for each sampler in \cref{fig:comparison} with and without distance marginalisation. \nessai requires a median of \nessaievaluations and \nessaievaluationsmarg likelihood evaluations to converge with and without distance marginalisation respectively and \dynesty requires \dynestyevaluations and \dynestyevaluationsmarg.  In contrast to \dynesty, sampling with \nessai is more efficient without distance marginalisation, we attribute this to a combination of the reparameterisation used for luminosity distance and the sampler settings we converged on for \nessai.

This, however, does not directly translate to the run-times for each sampler since they each have different additional computational costs associated with sampling. In \cref{fig:comparison} we show the total run-time for each sampler and when comparing the median run-times we observe that \nessai is \timeratio times faster than \dynesty without distance marginalisation and \timeratiomarg times faster with it. Additionally, we examine the proportion of the run-time spent on training and population and find that on average population-time accounts for approximately \populationtime of the total run-time and training-time accounts for a further \trainingtime. We also note that the cost of training and population does not depend on the cost of evaluating the likelihood, as such, the fraction of the total run-time will decrease as the cost of evaluating the likelihood cost increases.

\begin{figure}
    \centering
    \includegraphics[width=\figwidth]{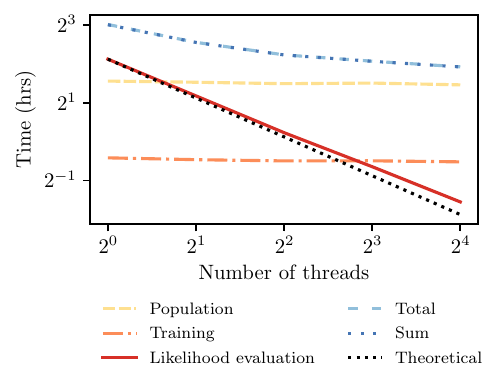}
    \caption{Comparison of the total time (in hours) spent on each stage of the algorithm for increasing number of threads for a single injection with a fixed noise seed. The time spent evaluating the likelihood decreases as the number of threads increases, the theoretical reduction is shown in black. The training and population stages remain approximately constant, as such act a lower bound on the minimum run-time. The sum of the time spent on likelihood evaluation, training and population is approximately equal to the total time spent sampling, indicating minimal overhead.}
    \label{fig:multithreading}
\end{figure}

For each injection we can also compare the posterior distributions produced by each sampler. These allow us to quickly identify discrepancies between samplers for specific injections or regions of the parameter space. However these differences are not easily quantified and the correlations between more than two parameters are not clearly represented. We show an example of such a comparison is \cref{app:corner}.

\subsection{Parallelisation of the likelihood computation}\label{sec:multiprocessing}

Our sampler is designed such that candidate live points are drawn simultaneously in the population stage. This allows for simple parallelisation of the likelihood computation since the pool of candidate live points can be distributed over a number of threads and likelihood values computed and stored until needed for the proposal stage. In \cref{fig:multithreading} we compare the run-time and time spent evaluating the likelihood for the same injection using increasing number of threads for the likelihood computation. We use an additional thread for the main sampling process. The time spent evaluating the likelihood is inversely proportional to the number of threads allocated although the overall run-time is not. With a single thread it accounts for \multievaltime of the total run-time and this decreases to \multievaltimemin when using 16 threads. As mentioned previously, there is a cost associated with populating stage and further smaller cost associated with training, for this injection these are \multipoptime and \multitraintime respectively. These remain approximately constant when increasing the number of threads available and act as a lower limit on the theoretical minimum run-time, this is shown in \cref{fig:multithreading}. The remaining $<1\%$ of the run-time is general overhead associated with running the sampler.

\subsection{Diagnostics}\label{sec:diagnostics}

As mentioned previously, there are various challenges when implementing a sampling algorithm. \nessai is designed to sample from within the constrained prior, in this case care must be taken to ensure that the prior is not over-constrained since this will lead to regions of parameter being under-sampled which in turn will bias the results. There are also specific problems that arise from the nature of the parameter space, such as multi-modality and correlations. We use a series of diagnostics to identify possible problems during sampling. In \cref{sec:tuning} we also discuss how some of these diagnostics can be used to tune the sampler settings described in \cref{sec:algorithm,app:sampling-settings}.

\begin{figure}
    \centering
    \includegraphics[width=\figwidth]{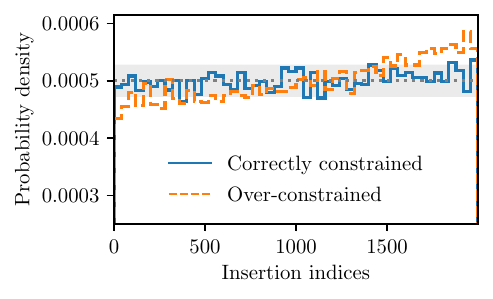}
    \caption{Example of the distribution of insertion indices for two nested sampling runs with 2000 live points. Uniformly distributed indices indicate no under- or over-constraining and deviations from uniformity indicate the opposite. The result with an orange dashed line shows over-constraining and the result with a solid blue line shows the correctly converged run. The shaded region indicates the 2-$\sigma$ errors on the expected distribution.}
    \label{fig:insertion_indices}
\end{figure}

\begin{figure}[ht]
    \centering
    \includegraphics[width=\figwidth]{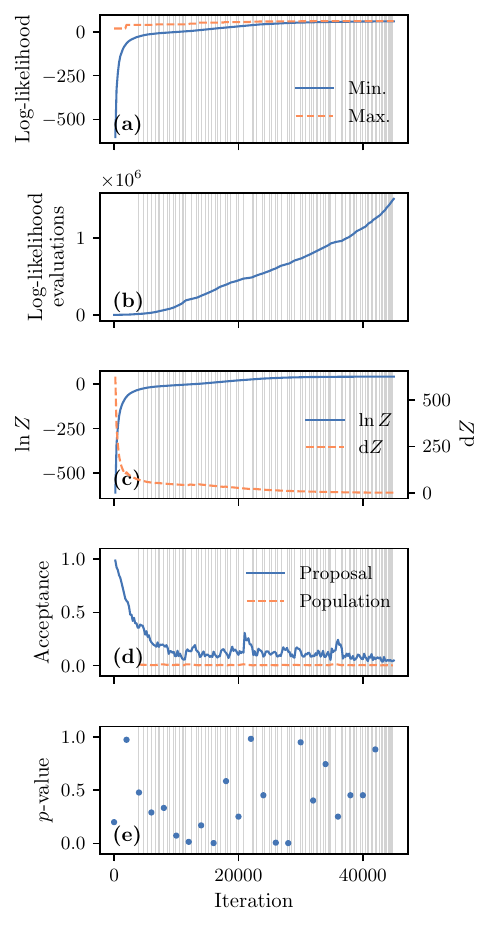}
    \caption{Example of the statistics that are tracked in our sampler as a function of sampling iteration: \textbf{(a)} minimum  (blue solid) and maximum (red dashed) log-likelihood, \textbf{(b)} cumulative number of likelihood evaluations, \textbf{(c)} log evidence $\log Z$ (blue solid) and fractional change in evidence $\text{d}Z$ (red dashed), \textbf{(d)} proposal (blue solid) and population (red dashed) acceptance and \textbf{(e)} $p$-value for cross-checks of every $K$ live points. The iterations at which the normalising flow is trained are indicated with vertical lines, for this injection these total \trainingits.}
    \label{fig:state}
\end{figure}

We use the cross-checks proposed in \cite{Fowlie:2020} as a heuristic for determining if the nested sampling algorithm has converged without over or under-constraining the posterior distributions. These checks rely on order statistics and the assumption that new live points should be inserted uniformly into the existing live points which allows for a $p$-value to be computed using the Kolomorgov-Smirnov statistic \cite{Smirnov:1948table} with the additional consideration that underlying distribution is discrete \cite{Arnold:2011npgof}. In \cref{fig:insertion_indices} we show an example of the distribution of the indices of newly inserted live points and in \cref{fig:state} we show the $p$-values computed every $K$ iterations. The histogram shows the final distribution of insertion indices for all the nested samples, this may not highlight specific problematic regions of the parameter space but if it is not uniform, it is a clear indication that the sampler is consistently over- or under-constraining. If the distribution of $p$-values in \cref{fig:state} is non-uniform then this is another clear indication of problems during sampling.

The acceptance is another important statistic to monitor during sampling since we aim to develop a more efficient sampler. There are two acceptances we can monitor in \nessai, the proposal acceptance and the population acceptance. The first has a direct effect on the number of likelihood evaluations whilst the second only affects the total run-time, both quantities are shown in \cref{fig:state}. This figure also highlights how periodically retraining the normalising flow leads to an increase in the proposal efficiency. It also shows how the population process is typically inefficient which explains why on average \populationtime percent of the total run-time is spent on the population stage.

We also track the minimum and maximum log-likelihoods, number of log-likelihood evaluations, log-evidence and fractional change in evidence. The combination of these statistics allows the user to quickly understand the current state of the sampler and identify potential issues such as plateaus in the likelihood space and regions which are inefficient to sample. The complete set of statistics is shown in \cref{fig:state}.

\subsection{Tuning \nessai}\label{sec:tuning}

\nessai includes various settings, a comprehensive list and description of each can be found in the documentation \cite{nessai-docs}. In practice we find that a small subset of the settings predominantly determine whether the algorithm converges without any bias. We use the validation method described in \cref{sec:validaton} and the diagnostics from \cref{sec:diagnostics} to understand how these settings affect convergence.

As expected, the number of live points $K$ is an important setting but it is even more crucial in \nessai since it limits the amount of training data available. We find that a minimum of 1000 live points is required and for more complex problems, such as gravitational-wave inference, at least 2000 live points should be used.

There are a large number of settings which relate to the complexity of the normalising flow. Whilst tuning the sampler we found that the number of coupling transformations greatly affected convergence. If too many transforms were used the algorithm was prone to over-constraining the posterior distribution. We attribute this to the complexity of the iso-likelihood contour learnt by the flow, if the flow has too many trainable parameters it can over-fit the distribution and exclude regions of the parameter space which should be sampled. At the other extreme, if the model is too simple then resulting contour can ``smooth'' fine details and more samples are drawn outside of the initial likelihood constraint. These will not be accepted and the sampling process is therefore less efficient. We use a similar logic for the number of neurons and layers in the neural network that parameterises the flow but we find that these parameters predominantly affect training time with a lesser effect on overall convergence. Another parameter that is important to consider is the batch size, during sampling the normalising flow can be training upwards of 100 times. Hence, a larger batch size is recommended since it can greatly reduce training time, we also recommend increasing the batch size when using reparameterisations that increase the amount of training data, such as the boundary inversion described in \cref{sec:algorithm,app:inversion}.

We note that the size of pool of new samples effects the efficiency of the algorithm and the total run-time. If the pool-size is small then the normalising flow is frequently retrained, in extreme case where the proposal is inefficient due to, for example, the complexity of the parameter space, then the normalising can be retrained multiple times during a single iteration. Conversely, if the pool-size is large then if the flow is force-ably retrained a number of points are discarded or, if the flow is only retrained once the pool is empty, then the rejection sampling becomes in-efficient since a large fraction of the potential new points will lie outside the likelihood bound. We instead opt to inversely scale the pool-size given the mean acceptance of the sampler since the last iteration the flow was trained. We recommend setting the base pool-size to the number of live points, only retraining the model when the pool is empty and setting the maximum pool-size to be ten times the base pool-size. We use these settings for the results in \cref{sec:results} and find that this results in a median of \npopulation training instances required to reach convergence.

As mentioned previously, approximately \populationtime of the run-time is spent on populating the pool of new samples. This is directly attributable to the efficiency of the rejection sampling required to ensure samples are distributed according to the prior. In \cref{sec:sampling} we propose two methods for drawing samples within the contour in the latent space, these produce uniformly and normally distributed samples respectively. In practice we find the two methods comparable in most cases with the exception of when the latent radius lies in the tail of the chi-distribution that corresponds to the latent prior $p_{\latent}$. In this case using the uniform distribution results in lower population and proposal acceptances which leads to longer run-times.

\section{Conclusions}\label{sec:conclusion}

We have proposed a novel method for sampling within a given iso-likelihood contour according to the prior that can be incorporated into the standard nested sampling algorithm. Our method employs normalising flows to learn the density of current set of live points which, once trained, allows us to produce points within the contour by sampling from a simple distribution and using rejection sampling. The use of normalising flows allows us to avoid using multiple bounding distributions and since new samples are independent of the previous samples we eliminate the need to use a random walk. We implement this proposal method in our sampler, \nessai, and conduct a series of tests to verify that it recovers the correct Bayesian posteriors and then compare our results to those obtained with another sampler to determine if our design does in fact result in a more efficient sampler.

We apply our sampler to 128 four second duration simulated signals from the coalescence of binary black hole systems sampled at 2048 Hz and we run two separate analyses, one with distance marginalisation and another without. The resulting P-P plots (\cref{fig:pp_plot}) show that our sampler more reliably recovers the posterior distributions with distance marginalisation than without, however both pass the P-P test. This indicates that our proposal method does not introduce any inherent biases.

We use \dynesty for the comparison, which has been shown to produce results consistent with those used in previous LVK analyses \cite{Romero-Shaw:2020}. We find that our sampler returns evidences consistent with \dynesty, which serves as further verification of our results. Since we aim to produce a more efficient sampler we also compare the likelihood evaluations required to reach convergence. When not using distance marginalisation we find that \nessai requires \nessaievaluations likelihood evaluations, \evaluationsratio times fewer than \dynesty. When distance marginalisation is enabled \nessai requires \nessaievaluationsmarg, which, whilst still \evaluationsratiomarg fewer than \dynesty, is more than with the marginalisation disabled. As such, we recommend using \nessai without distance marginalisation for gravitational-wave inference.

However, this reduction in likelihood evaluations does not relate directly to the total computation time because of the additional costs associated with sampling, which for \nessai are associated with training the normalising flow and populating the pool of new samples. We find that the fraction of the time spent of each stage changes when using distance marginalisation. Without the marginalisation, on average, \trainingtime of the total computation time is spent on training and a further \populationtime on population. When using distance marginalisation this changes to \trainingtimemarg spent on training and \populationtimemarg on population. We attribute the difference in population time to the efficiency of the rejection sampling, which is improved when including the reparameterisation for distance discussed in \cref{sec:gw_reparam}. We find that without distance marginalisation the median run-time for \nessai is \timeratio times faster than \dynesty. However when distance marginalisation is enabled we observe that, on average, \nessai is only \timeratiomarg times faster than \dynesty. This further reinforces our recommendation to use \nessai with distance marginalisation disabled.

We also show how our sampler can make use of parallelised likelihood functions by evaluating the likelihood of new live points during the population stage. We repeat the previous analysis for a single injection without distance marginalisation and parallelise the likelihood computation with increasing number of threads up to 16. We observe that the reduction time evaluating the likelihood does not quite match the theoretical values, indicating that there is a small overhead associated with it. This also highlights how the limiting factor is the time spent training the normalising flow and populating the pool of new live point.

To aid in diagnosing potential biases during sampling, we include a series of diagnostics in our sampler which allow us to easily identify under and over-constraining. These diagnostics also help to tune the sampling settings and highlight how periodically re-training the normalising flow during sampling prevents the proposal from becoming inefficient during sampling.

We find that our algorithm is susceptible to under-sampling regions of the parameter space which are close to the prior bounds. We consequently introduce the previously described reparameterisations to mitigate this and a series of diagnostics to aid in diagnosing biases and correctly tuning the settings. We aim address this in further work with changes to the design of the normalising flows we have used.

It is natural to compare this work to \cite{Gabbard:2019,Chua:2019,Green:2020a,Green:2020complete} which use variational autoeconders and normalising flows to produce posterior distributions. Our approach differs from these in that it requires no prior computation since training occurs during sampling and we do not introduce any assumptions about the data other than those necessary to apply a nested sampling algorithm. \nessai is therefore a drop-in replacement for existing sampling algorithms that does not require changes to existing pipelines.

In future work we aim to evaluate our sampler using more expensive waveform models including those for longer duration signals, such as those from binary neutron star of neutron star-black hole system, and models which include higher-order modes. We will also investigate the suitability of other types of normalising flow transforms, such as the spline based transforms from \cite{Durkan:2019nsf} and flows which allow for specifying a manifold \cite{Brehemer:2020manifold}. These changes could improve the efficiency of the population stage which is currently the slowest part of the algorithm. Another possible approach for reducing the cost of population is using alternative reparameterisations for parameters such as the spins magnitudes, which we observe to be two of the most challenging parameters to sample.

In summary, we have proposed a novel variation of the standard nested sampling algorithm that incorporates normalising flows specifically designed for inference with computationally expensive likelihood functions. We have applied our sampler to the problem of gravitational wave inference and shown that it consistently recovers the Bayesian posteriors distributions and evidences with \evaluationsratio times fewer total likelihood evaluations than \dynesty, another commonly used sampler, which translates to a \timeratio times reduction in computation time. Our sampler therefore serves as a more efficient drop-in replacement for existing samplers.

\begin{acknowledgments}

The authors gratefully acknowledge the Science and Technology Facilities Council of the United Kingdom. MJW is supported by the Science and Technology Facilities Council [2285031]. JV and CM are supported by the Science and Technology Research Council [ST/ L000946/1]. CM is also supported by the European Cooperation in Science and Technology (COST) action [CA17137]. The authors are grateful for computational resources provided by Cardiff University, and funded by an STFC grant supporting UK Involvement in the Operation of Advanced LIGO.

\textit{Software:} \nessai was initially developed using \cpnest \cite{Veitch:2021cpnest} with permission from the authors and still shares a similar interface and other core codes. \nessai is implemented in \python and uses \numpy \cite{numpy}, \scipy \cite{2020SciPy-NMeth}, \pandas \cite{reback2020pandas,mckinney-proc-scipy-2010}, \nflows \cite{nflows}, \pytorch \cite{Paszke:2019:pt}, \matplotlib \cite{Hunter:2007} and \seaborn \cite{waskom2020seaborn}. Gravitational wave injections were generated using \bilby and \bilbypipe \cite{Ashton:2019}. Figures were prepared using \matplotlib \cite{Hunter:2007}, \seaborn \cite{waskom2020seaborn}, \bilby \cite{Ashton:2019} and \corner \cite{corner}.

\end{acknowledgments}

\begin{widetext}

\appendix

\section{Loss function}\label{app:loss}

A normalising flow applies a mapping $f: \physical \to \latent$ conditioned on its parameters $\vec{\theta}$ which are typically the trainable parameters of a neural network. In this context the goal of training a normalising flow is to approximate a target distribution $p^{*}_{\physical}(x)$. The KL divergence between the target distribution and the distribution of the flow $p_{\physical}(x|\theta)$ can be written as \cite{Paramakarios:2019nfpmi}:

\begin{equation}
    \begin{split}
        \mathcal{L}(\vec{\theta})  = & D_{\textrm{KL}}[p^{*}_{\physical}(x) ||p_{\physical}(x|\vec{\theta}) ] \\
        = & - \mathbb{E}_{p^{*}_{\physical}(x)} [\ln p_{\physical}(x|\vec{\theta})] + \textrm{const.} \\
        = & - \mathbb{E}_{p^{*}_{\physical}(x)} \left[\ln p_{\latent}(f(x|\vec{\theta})) + \ln \left|\textrm{det} \frac{\partial f(x|\vec{\theta})}{\partial x^{T}} \right|\right] + \textrm{const.}.
    \end{split}
\end{equation}

Then, assuming the set of K samples used for training is drawn from $p^{*}_{\physical}(x)$, the expectation value can be approximated as:

\begin{equation}\label{eq:loss}
    \mathcal{L}(\vec{\theta}) \approx - \frac{1}{K} \sum_{k=1}^{K} \ln p_{\latent}(f(x|\vec{\theta})) + \ln \left|\textrm{det} \frac{\partial f(x|\vec{\theta})}{\partial x^{T}} \right|.
\end{equation}

\section{Boundary inversion}\label{app:inversion}

In \cref{sec:algorithm} we describe \textit{boundary inversion} which we introduce to avoid under-sampling regions which are close to the prior bounds. The user defines which parameters the inversion can be applied to and before training the sampler determines if it should be applied to each parameter using the following steps:

\begin{enumerate}
    \item {Compute the density of samples over the specified range and find the maximum value.}
    \item {Compute the fraction of the density that lies within the initial and final $m\%$ of the bounds, i.e. $[0, 0.1]$ and $[0.9, 1.0]$ if the parameter is defined on $[0, 1]$.}
    \item {Choose to apply inversion to bound with the highest density if it is at least $n\%$ of the maximum density and the density at the bound is non-zero.}
\end{enumerate}

From our testing the percentages $m$ and $n$ default to 10\% and 50\% respectively but can be changed by the user. We consider two methods for applying the inversion:

\begin{itemize}
    \item \textbf{duplication:} duplicate the set of points and apply the inversion to the duplicates,
    \item \textbf{splitting:} randomly select half of the points to apply the inversion to.
\end{itemize}

We find that duplication generally provides more consistent results but at the cost of the increasing the training time. As such we recommend using splitting when inversion is applied to more than two parameters.

\clearpage

\section{Gravitational-wave priors}\label{app:priors}

\begin{table*}[h]
\caption{Prior distributions used for each parameter for gravitational-wave parameter estimation. Their corresponding labels and the lower and upper bounds are included where applicable.}
\begin{tabularx}{\doublefigwidth}{Xccc} \toprule
    Parameters & Label & Prior & Bounds \\ \midrule
    Chirp mass & $\mathcal{M}$ & Uniform & $[25, 35] {\text{M}_{\odot}}$  \\
    Asymmetric mass ratio & $q$ & Uniform & $[0.125, 1.0]$ \\
    Luminosity distance & $d_{\text{L}}$ & Uniform in co-moving volume & $[100, 2000] {\text{Mpc}}$ \\
    Right ascension & $\alpha$ & Uniform & $[0, 2 \pi]$ \\
    Declination & $\delta$ & Cosine & - \\
    Reference time at geocentre & $t_{\text{c}}$ & Uniform & $[-0.1, 0.1]$ \\
    Inclination & $\theta_{\text{JN}}$ & Sine & - \\
    Polarisation & $\psi$ & Uniform & $[0, \pi]$ \\
    Phase & $\phi_{c}$ & Uniform & $[0, 2 \pi]$ \\
    Dimensionless spin magnitudes & $\chi_i$ & Uniform & $[0, 0.99]$ \\
    Spin tilt angles & $\theta_i$ & Sine & - \\
    Difference between the azimuthal angles of each spin vector relative to the orbital angular momentum & $\phi_{12}$ & Uniform & $[0, 2 \pi]$ \\
    Difference between the azimuthal angles of the  total and orbital angular momentum & $\phi_{\text{JL}}$ & Uniform & $[0, 2 \pi]$ \\
    \bottomrule
\end{tabularx}
\end{table*}

\section{\nessai sampling settings}\label{app:sampling-settings}

\begin{table}[h]
\caption{Settings used for \nessai for gravitational-wave inference. These are split into three categories: general settings which control aspects of the sampler such as the choice of latent prior or pool-size, flow hyper-parameters which determine the configuration of the normalising flow and flow training settings which control the training process. Different batch sizes were used for runs with and without distance marginalisation and this is shown in parentheses. For a complete description of each see the documentation \cite{nessai-docs}.}
\begin{tabular}{lP||lP||lP} \toprule
    \multicolumn{6}{c}{\nessai settings} \\ \midrule
    \multicolumn{2}{c||}{General settings} & \multicolumn{2}{c||}{Flow hyper-parameters} & \multicolumn{2}{c}{Flow training settings} \\ \midrule
    Training frequency & None & Coupling transformations & 6 & Optimiser & Adam \\
    Cooldown & 200 & Linear transformation & LU & Learning rate & 0.001 \\
    Base pool-size & 2000 & Network type & ResNet & Batch size & 2000 (4000) \\
    Update pool-size & True &  Layers per network & 2 & Max. epochs & 500\\
    Draw-size & 2000 & Neurons per layer & 32 & Patience & 50\\
    Train on empty & True & Activation & ReLU & \\
    Weights reset & False & Batch-Normalisation & Intra-transforms & \\
    Latent prior & Truncated Gaussian & & &\\
    Rescale & True & & &\\
    Update bounds & True & & &\\
    \bottomrule
\end{tabular}
\end{table}

\clearpage

\section{P-P tests for \dynesty}\label{app:dynesty}

\begin{figure}[h]
    \subfigure[With phase marginalisation]{\includegraphics[width=\figwidth]{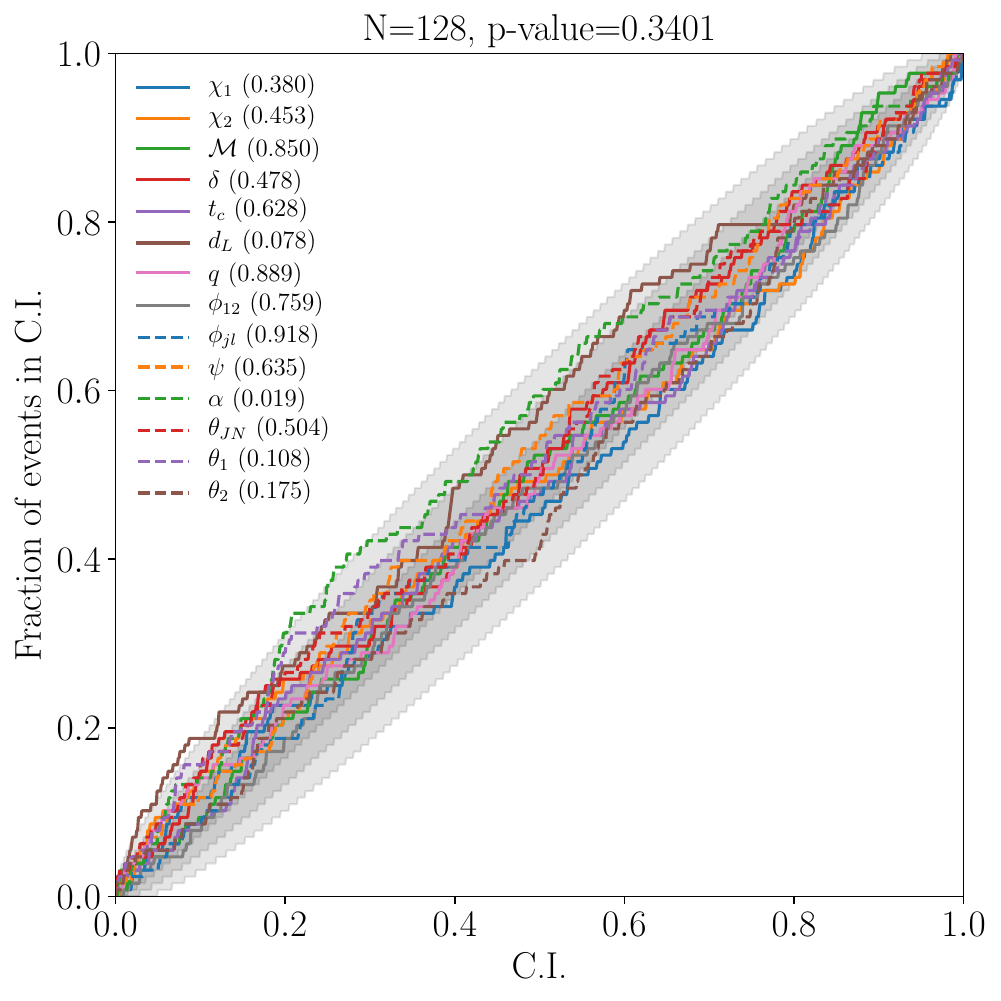}}
    \subfigure[With phase and distance marginalisation]{\includegraphics[width=\figwidth]{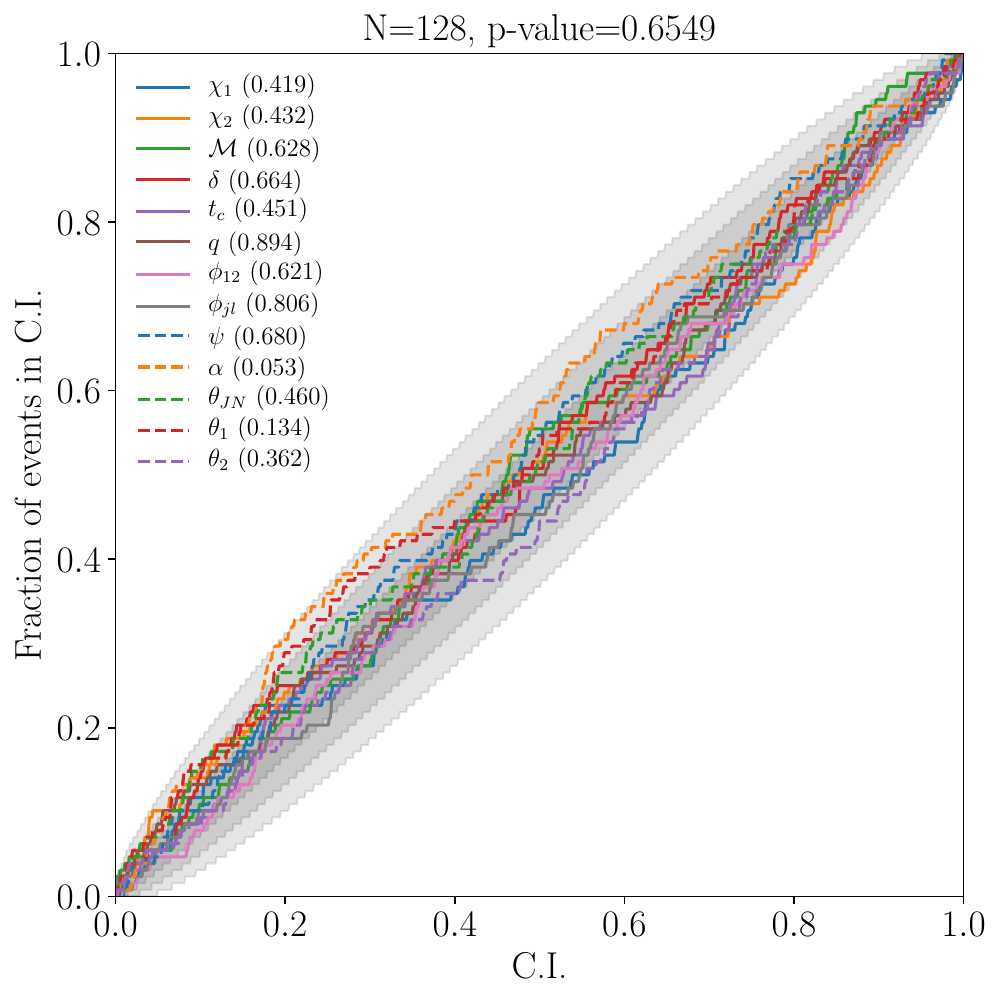}}
    \caption{Probability-probability (P-P) plot showing the confidence interval versus the fraction of the events within that confidence interval for the posterior distributions obtained using \dynesty for 128 simulated compact binary coalescence signals produced with \bilby and \bilbypipe. The 1-, 2- and 3-$\sigma$ confidence intervals are indicated by the shaded regions and $p$-values are shown for each of the parameters and the combined $p$-value is also shown. We use the settings described in \cite{Romero-Shaw:2020} with the exception of the number of live points which we increase to 2000.}
    \label{fig:pp_plot_dynesty}
\end{figure}

\clearpage

\section{Example corner plot}\label{app:corner}

\begin{figure}[h]
    \centering
    \includegraphics[width=\doublefigwidth]{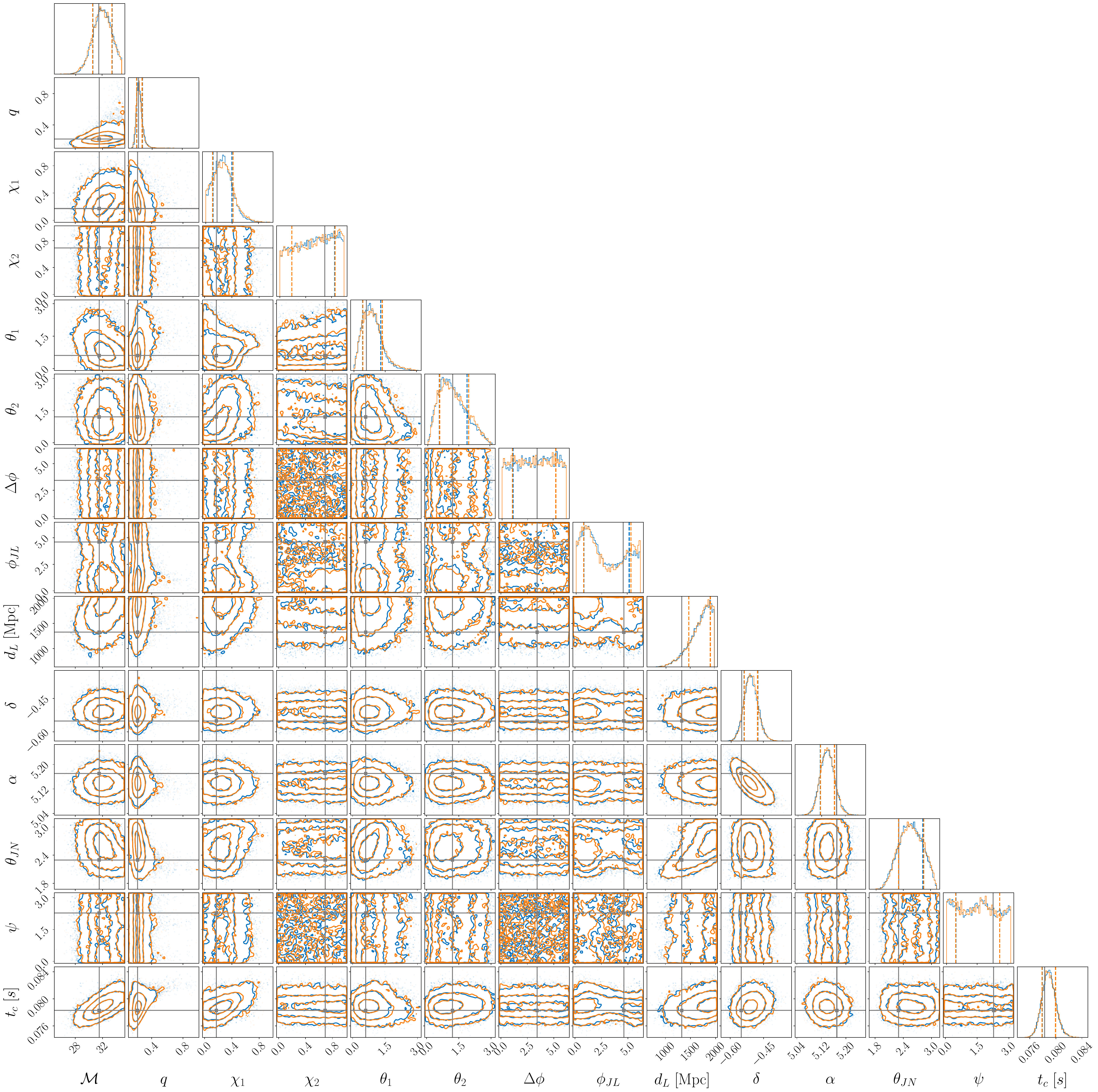}
    \caption{Corner plot comparing the posterior distributions produced with \dynesty (blue) and our sampler \nessai (orange) for an injection with an optimal network \ac{SNR} of \cornersnr. The phase is marginalised and remaining 14 parameters are shown, see \cref{app:priors} for details about the parameters. The injected value is indicated by the cross-hairs in each subplot and the respective 16\% and 84\% percentiles are also shown in the \protect\ndimensional{1} marginalised posteriors.}
    \label{fig:corner}
\end{figure}

\end{widetext}

\clearpage

\bibliography{ns_w_flows}

\end{document}